%% file: Formatting-Instructions-LaTeX-2026.tex
\documentclass[letterpaper]{article} % DO NOT CHANGE THIS
\usepackage{aaai2026}  % DO NOT CHANGE THIS
\usepackage{times}  % DO NOT CHANGE THIS
\usepackage{helvet}  % DO NOT CHANGE THIS
\usepackage{courier}  % DO NOT CHANGE THIS
\usepackage[hyphens]{url}  % DO NOT CHANGE THIS
\usepackage{graphicx} % DO NOT CHANGE THIS
\usepackage{natbib}  % DO NOT CHANGE THIS AND DO NOT ADD ANY OPTIONS TO IT
\usepackage{caption} % DO NOT CHANGE THIS AND DO NOT ADD ANY OPTIONS TO IT
\usepackage{algorithm}
\usepackage{algorithmic}
\usepackage{bibentry}
\usepackage{listings}
\usepackage{subfigure}
\usepackage{colortbl}
\usepackage{amsmath}
\usepackage{amsfonts}
\usepackage[table]{xcolor}
\usepackage{booktabs} %% Added by MJ
\usepackage{paralist}
\usepackage{multirow}
\usepackage{soul}  % for strikethrough \st
\usepackage{xspace}
\usepackage{pifont} 

\definecolor{mycolor}{HTML}{004C99}
\colorlet{royalblue}{mycolor!100}
\definecolor{mycolor}{HTML}{004C99}
\colorlet{royalblue}{mycolor!100}

\usepackage{newfloat}
\usepackage{listings}
\DeclareCaptionStyle{ruled}{labelfont=normalfont,labelsep=colon,strut=off} % DO NOT CHANGE THIS
\floatstyle{ruled}
\newfloat{listing}{tb}{lst}{}
\floatname{listing}{Listing}
\newcommand{\ksk}[1]{\textcolor{black}{#1}}

\newcommand{\uen}{\textsc{UeN}\xspace}
\newcommand{\yesmark}{\textcolor{green}{\ding{51}}}  % ✓ in blue
\newcommand{\nomark}{\textcolor{red}{\ding{55}}}   % ✗ in red

\newcommand{\bluecol}[1]{\textcolor{blue}{#1}}
\title{Real-World Challenges in Fake News Detection: Dealing with Posts by Cold Users}
\author{
    Sai Keerthana Karnam\textsuperscript{\rm 1},
    Abhirup Kundu\textsuperscript{\rm 1},
    Jashn Arora\textsuperscript{\rm 2},
    Manish Jain\textsuperscript{\rm 2},
    Animesh Mukherjee\textsuperscript{\rm 1}
}
\affiliations{
    \textsuperscript{\rm 1}Indian Institute of Technology Kharagpur \\
    \textsuperscript{\rm 2}Google DeepMind\\
    \{saikeerthanakarnam.24@kgpian, animeshm@cse\}.iitkgp.ac.in, 
    \{arorajashn, manishjn\}@google.com. \\
}
\usepackage{bibentry}
\begin{document}

\maketitle

\begin{abstract}
Social media serves as a primary source of information in the current digital era. Many people consume a vast range of information in a very short span, yet, amidst the stream of genuine information, fake news and rumors continue to spread. The need for effective detection models is becoming increasingly critical. Past user behavior and user engagement on a post are strong signals that SOTA approaches leverage for fake news detection and other post classification tasks. However, these approaches lean too heavily on knowing this past behavior, and thus suffer from a \textit{cold user problem}, or users that are new or have minimal footprint on the platform. In this paper, we make three core contributions. We first establish the value of user behavior, both content and user-user interactions, in the task of fake news and rumor detection. We then establish the extensive prevalence of \textit{cold users} in the real-world datasets, and show the need for newer algorithms considering cold users. We next propose a novel socially-aware context representation scheme – \textsc{User Evidence Network} (\uen) – to detect the spread of misinformation and unverified information while efficiently navigating this cold user challenge. We introduce techniques that approximate missing / absent behavior data of a new user from existing users' interactions. By carefully addressing the cold user challenge, our work provides robust approaches targeting fake news and rumor detection for real-world platforms. 
\end{abstract}
% Uncomment the following to link to your code, datasets, an extended version or similar.
%
\noindent\rule{\columnwidth}{0.4pt}
%\noindent\textsuperscript{*}These authors contributed equally to this work.

\bluecol{The code is available at \url{https://github.com/saikeerthana00/UEN}. This work has been accepted at ICWSM 2026 as a full paper.}
 
% \begin{links}
%     \link{Code}{https://github.com/saikeerthana00/UEN}
% \end{links}
\input{Introduction.tex}

\input{Related.tex}

\input{Dataset.tex}

\input{Problem_formulation}
\input{UEN_framework.tex}

\input{Results.tex}

\input{Conclusion.tex}

\section{ Ethics statement}
This study does not aim to identify or trace individual users involved in the dissemination of fake news. Our intent is not to harm any individuals or target specific communities. All experiments were conducted on publicly available or previously published datasets. The primary objective of this research is to enhance model generalization, particularly for scenarios involving cold-start users.
\bibliography{mybibfile}
\end{document}

%% file: Introduction.tex
\section{Introduction}

In the present digital age, social media platforms have emerged as a primary means by which people acquire information. Every day, consumers are exposed to a diverse range of content, from accurate news to unverified and misleading information. The continued rapid proliferation of fake news on platforms like Meta \citep{articlefacebook,article_facebook_2}, X (formerly Twitter) \citep{article_twitter, article_twitter_2}, and Reddit \citep{10.1145/3342220.3343662} highlights the critical need for effective detection techniques. 

\noindent Recently, \citet{gong2023fakenewsdetectiongraphbased} presented a survey on fake news detection models and highlighted the effectiveness of graph neural networks (GNNs) in identifying fake news, because GNNs can simulate the social contexts and propagation patterns of information. Building upon GNNs, \citet{article,yuan2019jointlyembeddinglocalglobal,bian2020rumordetectionsocialmedia} all present approaches that integrate both \textit{local} and \textit{global features} of the network. Local features include individual user interactions and immediate comment threads, while global features capture broader patterns across the network, such as overall communication trends and influence metrics.\\
\noindent\textbf{Leveraging user behavior}: \citet{Yang2012AutomaticDO} and \citet{Castillo2011InformationCO} demonstrated the significance of incorporating user profile features, such as age, number of tweets, and followers, in enhancing the accuracy of fake news detection. However, this information is often incomplete, unavailable, or inaccurate, and metadata features can be biased. Further such information may lead to compromise of user privacy. To address these issues, we propose the \textsc{User Evidence Network} (\uen) based framework which relies solely on the commenting and replying patterns of users to efficiently capture their behavioral traces.\\ 
\noindent\textbf{Cold users}: While past information on user behavior can bring in steady benefits to the GNN based fake news detection task \cite{10.1145/3289600.3290994}, a major shortcoming is the absence of such information for a section of users who have just joined the platform or have done very little interactions so far. In this paper we call them the \textit{cold users}. The lack of digital footprint in terms of prior posting and commenting patterns for these users can severely handicap the workings of the model and adversely affect the overall performance. To overcome this limitation, our work proposes novel heuristics to create (approximate) feature representations for the cold users by mapping their behavioral pattern to the existing users. This approach aims to improve the robustness of fake news detection models by mitigating the limitations associated with unavailability of traditional user profile and behavior features.\\%\mj{user profile ?}\\
% \mj{for next statement - as a style, we are telling the solution (`introduce heuristics') \textit{before} describing the problem (`cold start'). This is an important enough problem for us to describe up-front.} 
% \mj{We should talk about how approaches broadly suffer from lack of data across the dimensions we listed earlier (content/user interaction/behavior). Content is understood e.g. language models, but user interaction / user behavior cannot be understood for \textit{new users}. Then a statement on the prevalence of this  `cold start` problem, and then motivate the need to address this as a first principle. Then our approach - and we want to make a stronger statement than `introducing heuristics`. Talk about our evals, generalization.}
\noindent\textbf{Temporal ordering}: The interactions around a fake news post keep evolving in social media platforms. Thus, in the real world, the entire temporal context about a post might not be always available to ascertain whether that post is fake or not. The authors in \citet{gong2023fakenewsdetectiongraphbased} highlights a key limitation in SOTA models where the complete future information of all posts are assumed to be available for detection. To maintain the realistic rigor in our work we take non-overlapping training and test snapshots ordered in time and thereby do not rely on all the future data of a post. Our dataset creation approach mimics an actual implementation on a social media platform - models are developed on historical data, get deployed and provide value by classifying new unseen posts/comments. 
%evals maintain realistic rigor and 
%In a review of GNN models for fake news detection, \citet{gong2023fakenewsdetectiongraphbased} highlights a key limitation in current experiments: most existing methods train the models over the entire time span of the samples, giving them with a complete information to identify whether or not the news is fake. This approach may not reflect real-world scenarios where complete data is often unavailable, leading to overestimation of model performance. To address this issue, the 
%\une{} framework focuses on capturing the dynamic and evolving nature of information spread. We solve this issue by sorting the dataset by timestamps and then dividing it into training and testing groups. 
% \mj{This sounds a bit simplistic for the abstract - we can simply state that our evals maintain realistic rigor and take snapshots in time and don't rely on all the future data of a post - something like that. Suggestion is to describe everything in the problem -> intuition -> solution framework and not directly begin with the solution.}

\begin{table*}[ht]
\scriptsize
\centering
\begin{tabular}{p{3.5cm}|p{1.8cm}|p{2.6cm}|p{3.3cm}|p{3cm}}
\hline
\textbf{Paper} & \textbf{Content-based} & \textbf{Propagation patterns} & \textbf{Historical user behaviour} & \textbf{Handles cold users} \\

\hline 
\citet{rashkin2017truth}, \citet{miyazaki2023fake}, \citet{nakamura2020rfakedditnewmultimodalbenchmark} & \yesmark & \nomark & \nomark & \nomark \\
\hline

\citet{han2020graph}, \citet{bian2020rumordetectionsocialmedia} , \citet{nakamura2020rfakedditnewmultimodalbenchmark}, \citet{xu2023rumor}, \citet{wei2024transferring}  & \yesmark & \yesmark & \nomark & \nomark \\
\hline

\citet{yuan2019jointly}, \citet{sun2023hg}, \citet{su2023hy}  & \yesmark & \yesmark & \yesmark & \nomark \\
\hline

\citet{10447588} & \yesmark & \yesmark & \nomark & \nomark  (Handles cold start problem but doesn't include user features)\\
\hline

Our work  & \yesmark & \yesmark & \yesmark & \yesmark \\
\hline

\end{tabular}
\caption{\footnotesize Prior works leverage content features, model propagation patterns, and incorporate historical user behaviour, but lack the ability to handle cold users. In contrast, our work addresses all four aspects, with a particular emphasis on cold user generalization.}

\label{tab:related_work}
\end{table*}

The main contributions of this work are as follows.
\begin{compactitem}
%\item We capture user behavior as a valuable indicator for fake news detection by constructing a global interaction-based graph that collects user information from post-user interactions via commenting and replying behaviors. We do not rely on other user attributes like followers, likes, gender, etc. 
% \mj{this is a reviewer question waiting to be asked - ablation study.}
\item We propose the \uen framework that makes comprehensive use of the content of the source post, comments or reactions, comment pattern or tweeting pattern, and user information obtained from a global interaction-based graph to train GNN model variants to detect fake news or rumor.
\item To improve the model's ability to generalize for samples involving cold users, we suggest multiple strategies for extracting feature representations from the global interaction-based graph. Even in the absence of prior data, we generate (approximate) meaningful feature representations by mapping the cold users to some of the existing users employing an array of novel heuristics.
\item We show that incorporation of user evidence based features along with the textual features of the source post improves the accuracy of the prediction by $6-8\%$ for the different GNN variants. Further improvements are obtained in macro-F1 when feature representation of the cold users are acquired based on our proposed heuristics. Remarkably, the performance for the test users who have absolutely no footprint in the training set (`perfectly cold users') increases by around $9-10\%$ (macro-F1) when our proposed heuristics are employed. 

 % \mj{comments in same vein as before about lacking the higher altitude story arc.}
\end{compactitem}

%% file: Related.tex
\section{Related work}
\label{sec:related}

The early methods for fake news detection relied mainly on content-based features, such as linguistic cues and sentiment analysis, to identify misinformation \cite{rashkin2017truth}. \citet{miyazaki2023fake} used embedding features from sentenceBERT to train fake news classifiers using SVM and Random Forest, and to also finetune language models like BERT for the same task. \citet{nakamura2020rfakedditnewmultimodalbenchmark} integrated linguistic features using BERT embeddings, and visual features using InferSent and used a combination of both of these to create a fake news classifier.
%use BERT to embed the linguistic features and InferSent to embed the visual features of the content to classify fake news. 
%However, these content based techniques often proved insufficient, as fake news creators frequently employ sophisticated strategies to mimic genuine content.\\
To improve detection accuracy, recent approaches have incorporated social context into their models. 
They tend to leverage post content features plus the content of the comments, the user profile descriptors (user's self-description, age of account, verified status), user connectivity (number of friends and followers), and structural and dispersion features of the post (post-comment propagation tree).
Graph neural networks (GNNs) have emerged as a popular tool to model all these features of the post comment tree to build fake news classifiers. \citet{han2020graph} trained GNNs incrementally for this task using the user profile and tweet timeline along with the structural features. On the other hand, \citet{bian2020rumordetectionsocialmedia} 
proposed a bidirectional graph model, 
Bi-Directional GCN (Bi-GCN), and explored patterns of deep propagation and the structures of wide
dispersion in rumor detection by operating on both the top-down
and bottom-up propagation of rumors. \citet{xu2023rumor} proposed  Hierarchically Aggregated GNN (HAGNN) and focused on 
different granularities of high-level representations of text content and fused the rumor propagation structure. \citet{wei2024transferring} addressed the challenge of fake news detection for new posts by learning how to transfer structural knowledge learnt from existing propagation trees and classifying new posts with just their content. 
% \citet{10447588} proposed a method to address the cold start problem in propagation by transferring structural information.
Specifically, they utilised both propagation and content features during training, but tested on samples that contain only content features. \\
Much of this body of work focuses on the `local' social context of one particular post, and discards the `global' social media network when addressing the task of detecting fake news. Global signals include a user's post/comment history, history of engagement on the platform, and the set of other users they interact with, and these global signals have significant value when considering the task of fake news detection. \citet{sun2023hg} proposed a joint learning model named HG-SL, which is
%which is blind to news content and user identity, but 
capable of catching the differences between true and fake
news in the early stages of propagation through global and
local user spreading behavior, 
but is blind to the content.
\citet{yuan2019jointly} presented a novel global-local attention network
(GLAN) for rumor detection, which jointly encodes the local
semantic and global structural information by modelling the global relationships among all source tweets, retweets, and users as a heterogeneous graph to capture the rich structural information for rumor detection. \citet{su2023hy} proposed
constructing an attributed hypergraph to represent non-textual
and high-order relations for user participation in news spreading. %Their method captures semantic information from news content, credibility information from involved users, and high-order correlations between news and users to learn distinctive embeddings for fake news detection. 
However, all these approaches assume the availability of historical user data, making them ineffective in handling cold users, or users with minimal footprint on the given social media platform.

Several studies have explored the cold start problem in recommendation systems \ksk{\citet{10.1007/s00500-019-04588-x}, online social networks \citet{10.1145/3409108}} and other domains \citet{park2009pairwise,schein2002methods}. These methods typically employ techniques like content-based filtering or collaborative filtering with side information to make predictions for new users. 
However, the cold user challenge in fake news detection presents unique complexities, as the users do not reveal whether a news is fake or not themselves. 
%%INTENT HERE is to suggest that in a recommendation system like Netflix, even the cold users reveal their preferences when prompted, or when they make a search/selection. Here, the users themselves don't reveal any information and, at times, are trying to fool the system and pass fake news as good.
% as the absence of historical behavior data can significantly impact the model's ability to assess the credibility of users and their contributions.
Table \ref{tab:related_work} summarizes prior works and the types of features they utilize: (a) Content – such as titles or news content; (b) Propagation patterns – including tweet-retweet or post-comment structures; (c) Historical user behavior – such as user-user interactions and posting habits; and (d) Cold user handling – the model's ability to generalize to users with limited (or no) historical data. Our work builds upon previous research in fake news detection and cold-start problems. We propose a novel approach that explicitly addresses the cold user challenge by leveraging existing user-user interactions to approximate the missing user's behavior data. By carefully integrating social context and user behavior, our model provides robust and practical solutions for fake news detection in real-world scenarios.

\begin{table}[t]
\centering
\footnotesize

%\resizebox{.95\columnwidth}{!}{
\resizebox{\columnwidth}{!}{
\begin{tabular} { l|r|r } 
    \toprule
    \textbf{Statistic} & \textbf{Fakeddit dataset (text)} &  \textbf{Gossipcop dataset (text)}\\
    \midrule     
    Total \#samples & 1,063,106 & 5,464\\
    Text \#samples & 971,806 & 5,464\\
    Text \#samples ($|$comments$|$ $\ge1$) & 583,305 & 5,464\\
    Comments & 9,291,298 & 294,288\\
    \midrule
    Fake \#samples &  179,961 & 2,732\\
    True \#samples &  402,753 & 2,732\\
    \midrule
    Unique \#users & 1,592,037 & 76,356\\
    Cold \#users & 430,439 & 14,273\\
    \midrule
    Training set size & 407,760 & 3,824 \\
    Training set timespan  & 1/6/2008 - 3/21/2019 & 5/13/2008 - 2/16/2018
\\ 
    \midrule
    Validation set size & 58,315 & 546 \\
    Validation set timespan & 3/21/2019 - 6/7/2019 & 2/16/2018 - 4/13/2018 \\
    \midrule
    Testing set size & 116,639 & 1,094 \\        
    Testing set timespan & 6/7/2019 - 10/23/2019 &  4/13/2018 - 12/17/2018\\ 
    \bottomrule
\end{tabular}
}
\caption{\footnotesize Dataset statistics: The time period between the training set and testing set do not overlap ensuring that there is no data leakage.}
\label{table1}
\end{table}

% \textbf{Content based methods}

% \textbf{Content + local network based methods}

% \textbf{Content + local network + Global network (user behavior) methods}

% \textbf{Cold user/start}

%% file: Dataset.tex
\section{Dataset}
For our experiments, we use the \textbf{Fakeddit}~\cite{nakamura2020rfakedditnewmultimodalbenchmark} and \textbf{Gossipcop}\footnote{\url{https://github.com/safe-graph/GNN-FakeNews}}datasets.

\noindent \textbf{Fakeddit} is derived from Reddit and consists of over 1 million submissions from 22 subreddits over a 10 year span. Each submission has a submission title and image, an author, comments made by users who engaged with the submission, and other metadata like the comments count and username of the author. For the classification, we considered a binary label, which represents whether the submission is \texttt{fake} or \texttt{true}. The original paper addresses the classification of multimodal samples, thus considering submissions with text, images, or a combination of both. In this paper, we will exclude submissions that have only images as the content since this number is substantially less. In our framework, we incorporate comments associated with each submission. Therefore, we consider samples with at least one comment. As depicted in Table \ref{table1}, we use $583,305$ samples for our experiments. 
% The choice of this dataset is motivated by the fact that it is one of the latest and largest benchmark datasets for fake news detection. While there are others like the PHEME dataset, they are quite dated and small in size. Further, there is no easy way to expand such datasets due to the prohibitive cost of the X API.

\noindent \textbf{Gossipcop} dataset comprises source news articles along with associated tweets and retweets from X (formerly Twitter) conversations. Each tweet or retweet includes its text and the author information. Unlike the Fakeddit dataset, this does not have a user associated with the source post; therefore, we assume all source news items originate from a single, common user. For classification, we adopt a binary label indicating whether the source news is \texttt{fake} or \texttt{true}. Since only tweet IDs are publicly available and accessing tweet content via the X API is cost-prohibitive, we rely on pretrained BERT embeddings for tweet representations.
% In summary, each of our samples includes the submission or post title, author, comments associated with the submission, and the corresponding users. Even for comments, we include only text while omitting the images. In this paper, we refer to the terms submission and post interchangeably. We sort the data samples according to the timestamp and consider the first $70\%$ as the training set, the next $10\%$ as the validation set, and the remaining $20\%$ as the test set. This ensures that the test samples are completely unknown and reflect real-world scenarios. We define cold/unknown users as those who are either non-existent or not present in the training samples. Table \ref{table1} represents the statistics of the dataset.

In summary, we utilize two datasets - Fakeddit and Gossipcop for our analysis. Each of our samples includes the submission or source news, author, comments or reactions associated with the submission or tweet-retweet, and the corresponding users. For comments (in Fakeddit), we include only text while omitting the images. In this paper, we use the terms submission (from Fakeddit) and source news (from Gossipcop) and post interchangeably, as well as comments/replies (from Fakeddit) and tweets/retweets (from Gossipcop), as they are treated similarly in our pipeline. We sort the data samples according to the timestamp and consider the first $70\%$ as the training set, the next $10\%$ as the validation set, and the remaining $20\%$ as the test set. This ensures that the test samples are completely unknown and reflect real-world scenarios. We define cold/unknown users as those who are either non-existent or not present in the training samples. Table \ref{table1} notes the statistics of the dataset.

\begin{figure*}[t]
    \centering
    \subfigure[Pool of post-comment chains.]{
     \includegraphics[height=4cm,width=0.45\textwidth]{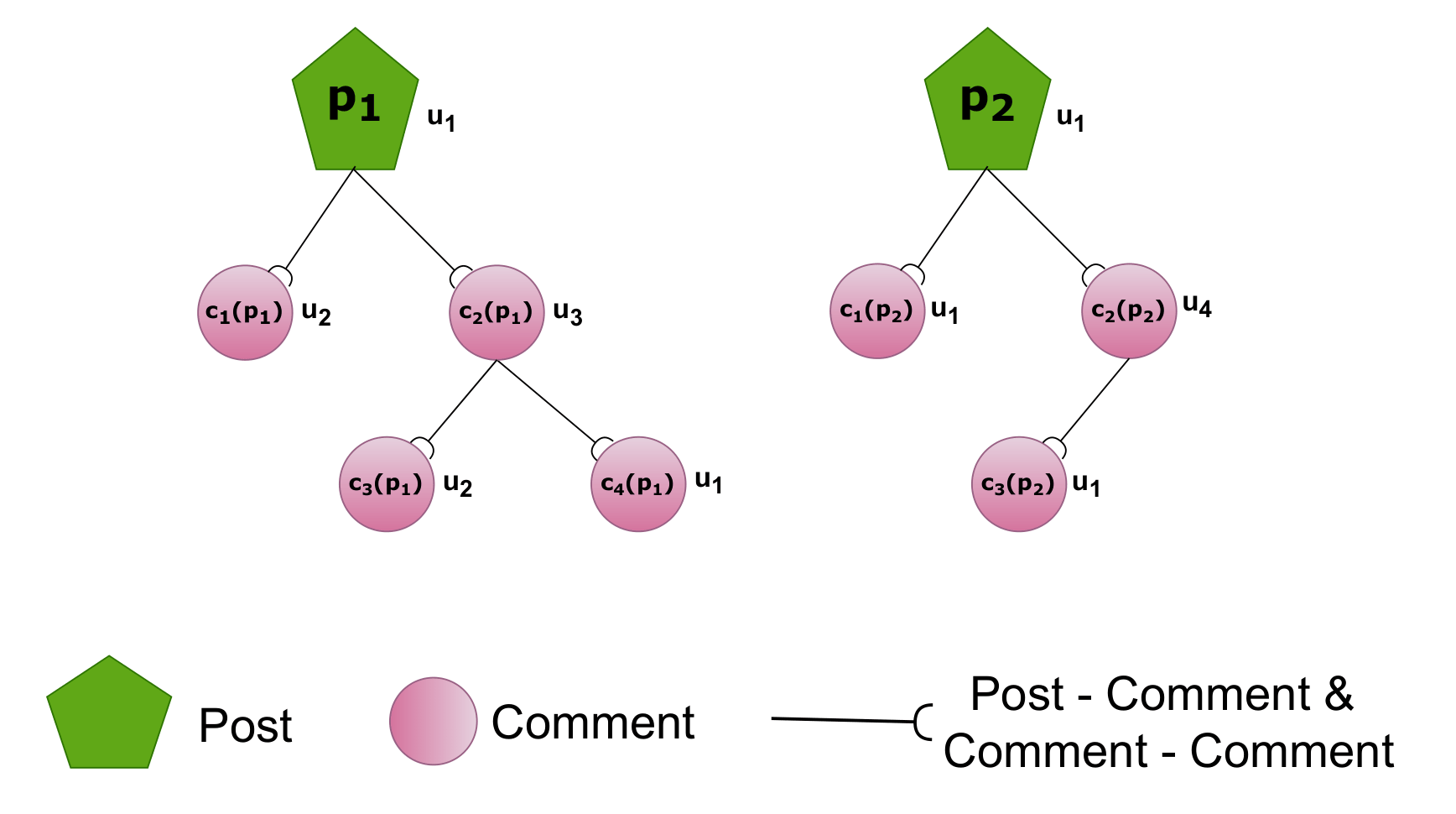}
    }
    \hspace{0.05\textwidth}
    \subfigure[Graph depicting the user-user interactions.]{
    \includegraphics[height=4cm,width=0.30\textwidth]{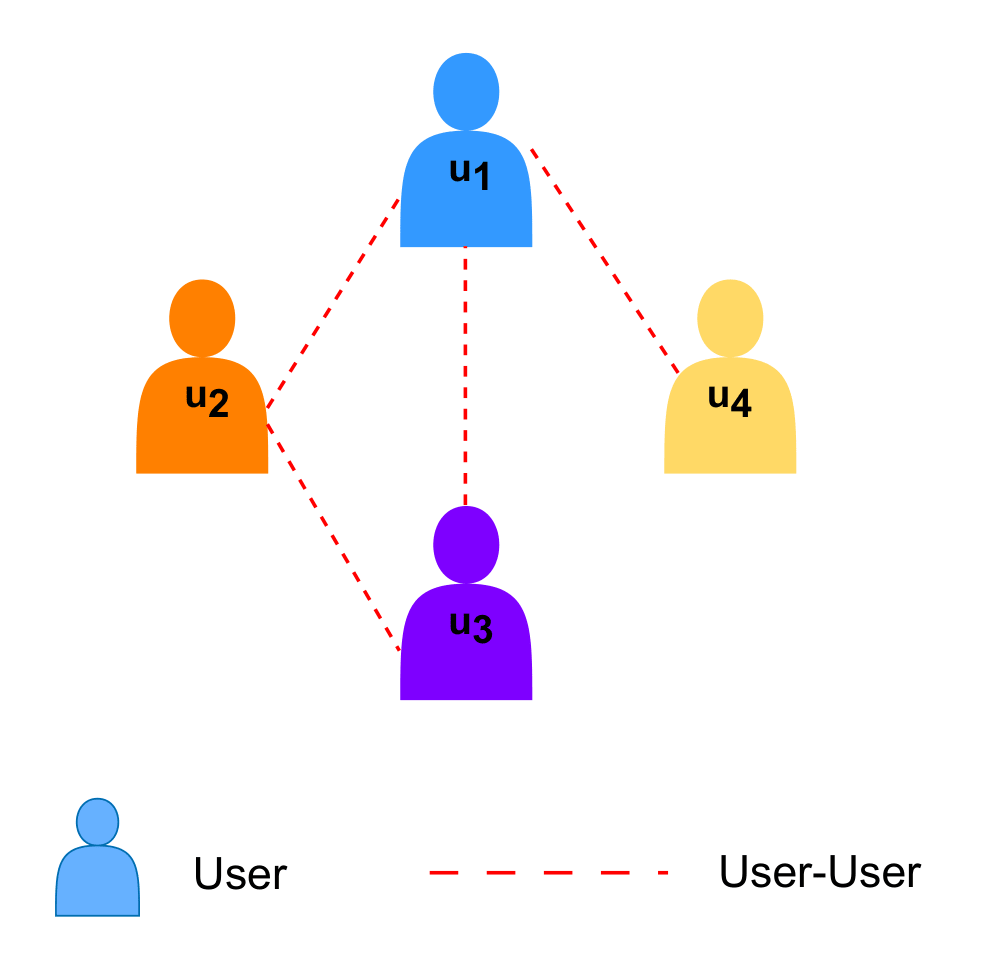}
    }
    \caption{Graph representation of the social context.}
    \label{pf}
\end{figure*}

\begin{figure*}[ht]
\centering
\includegraphics[height=11.8cm, width=0.80\textwidth]{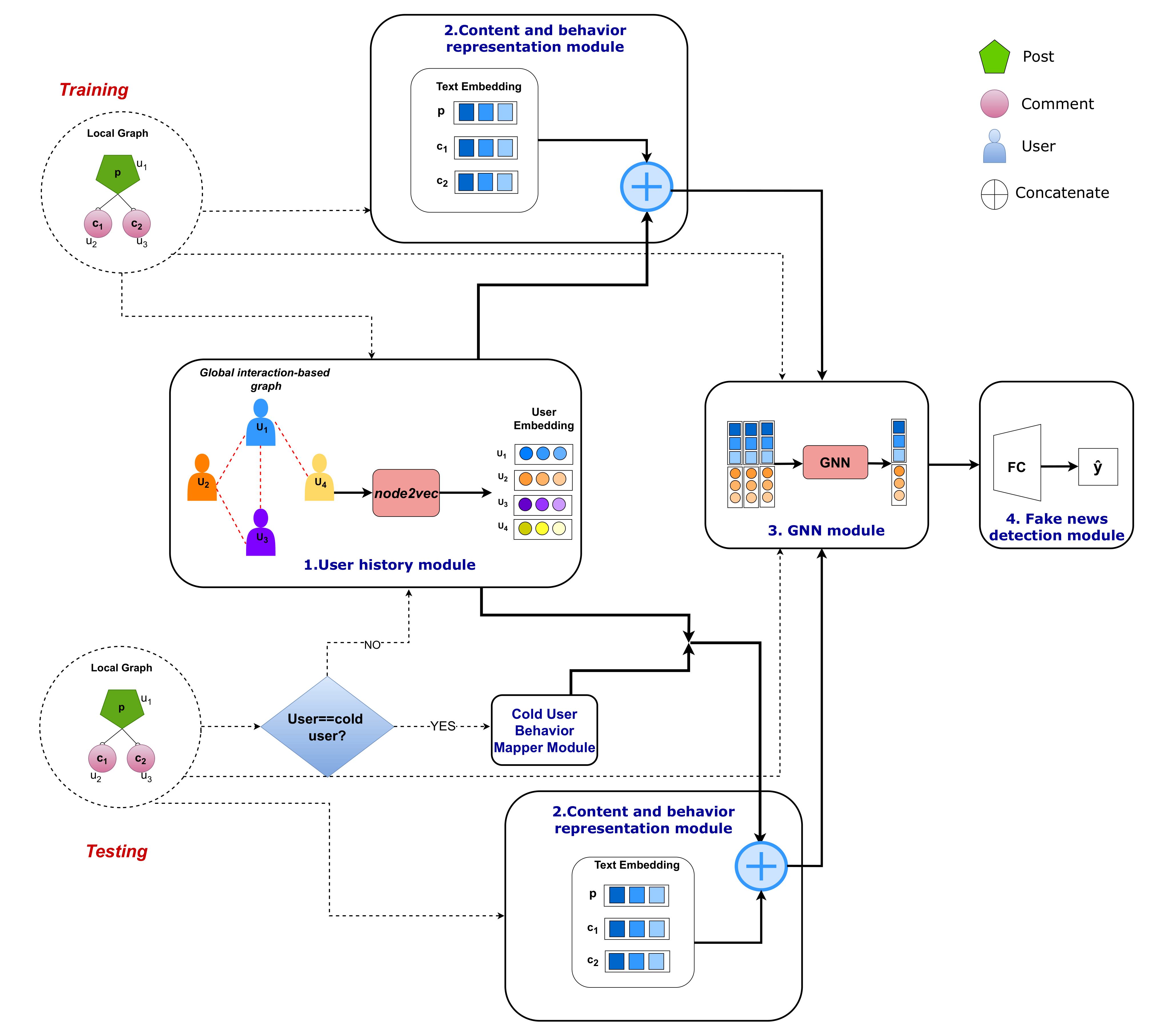} % Reduce the figure size so that it is slightly narrower than the column. Don't use precise values for figure width.This setup will avoid overfull boxes.
\caption{\footnotesize The overall architecture of the \uen framework. For \textbf{training}, first the \textit{global interaction-based graph} is constructed and trained to generate user representation in the first module. These embeddings are passed to the second module, which captures content and user behavior features. These features are fed to the third module to obtain a robust graph representation, which is ultimately classified in the fourth module. For \textbf{testing}, each user in the sample is first checked to determine if they are a cold user. Cold user representation is obtained from the cold user behavior mapper module (shown in Figure \ref{fig:main_heur}); while other users' representation is obtained from the first module. Remaining steps follow the training phase.}
\label{fig1}
\end{figure*}

%% file: Problem_formulation.tex
\section{Problem formulation}
\label{sec:problem}
We use the following notations in the rest of the paper. 
\begin{compactenum}
\item $P = \{p_1,p_2, \dots\}$ represents posts which have textual content and also have at least one textual comment associated with them.
\item $C(p_i) = \{c_1(p_i), c_2(p_i), \dots, c_n(p_i)\}$ represents the comments under the post $p_i$. Here $n$ represents the number of comments under the post $p_i$.
\item $E_i$ is the list of edges which depicts the local relations as follows.
\begin{compactenum}

 \item \textit{Post-comment edges} $(p_i,c_j(p_i))$: Represents that a comment $c_j(p_i)$ is made on the post $p_i$.   
 \item \textit{Comment-comment edges} $(c_j(p_i),c_k(p_i))$: Represents that a comment $c_k(p_i)$ is made in reply to the comment $c_j(p_i)$. 
\end{compactenum}
Both types of edges are treated equally in the pipeline.
\item $U_i$ is the set of social media users who either published the post $p_i$ or made a comment on the post $p_i$.
\item $y_i=\{0,1\}$ represents the ground truth assigned to source post $p_i$ with 0 indicating that the post is fake and 1 indicating that the post is true. Similarly, $ \overline y_i $ represents the predicted outcome.
\item $S_{\text{Train}}=\{s_1,s_2,s_3 \dots\}$ is the list of training data samples. Each data sample is denoted as $(p_i,C(p_i),E_i,U_i,y_i)$.
\item $S_{\text{Test}} = \{s_1,s_2,s_3 \dots\}$ is the list of test data samples. Each data sample is denoted as $(p_i,C(p_i),E_i,U_i)$.
\newcommand{\wbigcup}{\mathop{\widetilde{\bigcup}}\displaylimits}
\item $U_G$ represents the global set of social media users.
\small
\begin{equation}
\label{equa:user}
    U_G = \bigcup_{s_i \in S_{\text{Train}}} U_i
\end{equation}
\normalsize
\item $E_G$ is the list of edges which depicts the global relations between users $U_G$. Edge $(u_i,u_j)$ exists if there is an interaction between $u_i$ and $u_j$, where $u_i,u_j \in U_G$. The interactions could be of the following types.  
\begin{compactenum}
    \item User $u_i$ commented on a post that was created by $u_j$.
    \item User $u_i$ replied to a comment made by $u_j$.
\end{compactenum}

\end{compactenum}
Our objective is to train a classifier capable of assigning a label 0 or 1 to the data samples $S_{train}$ and $S_{test}$. Figure~\ref{pf} represents the comprehensive details and interactions used to execute this task.

% \mj{add visual of a post and connect to proposed formulation. Also worthwhile to point out as to what we are not doing - which is using some external data repository / world knowledge to compare news as fake or not.} 

%% file: UEN_framework.tex
\section{The \uen framework}
\label{sec:model}

The proposed \uen framework consists of five main components. First, the \textbf{User history module} uses the global relations mentioned in the problem formulation section to generate a feature representation for each user. Next, the \textbf{Content and behavior representation module} generates detailed representation for posts and comments by concatenating features from textual content and user behavior. These combined features along with the local relations are then fed to \textbf{GNN module} to find an enhanced representation of the data sample. Then, this representation is passed through the \textbf{Fake news detection module} to determine whether the post is fake. For testing, we use the \textbf{Cold user behavior mapper module} to find the representation for cold users. Figure \ref{fig1} represents the overall architecture of the model. 

\subsection{User history module}
We construct an undirected \textit{global interaction-based graph} $G(U_G,E_G)$ where nodes are denoted by $U_G$ and $E_G$ represents the set of edges. We define \( \mathbf{u_i} \) $\in \mathbb{R}^{d_1}$ as a $d_1-$dimensional embedding representation for user $u_i \in U_G$. We use the \texttt{node2vec}~\cite{grover2016node2vecscalablefeaturelearning} model, which computes the node embedding by sampling its neighborhood and then minimizing the proximity loss function. Using this module, we represent user embedding, which captures historical evidence and global interactions between the users.\\
\noindent For this module, we consider nodes $U_G$ as the set of users belonging to training data samples as shown in equation \ref{equa:user} and $E_G$ as the set of edges representing the interactions between the users in $U_G$. We train the global graph using the \texttt{node2vec}~\cite{grover2016node2vecscalablefeaturelearning} model and generate the user embeddings of dimension 128 ($d_1$).
\subsection{Content and behavior representation module} 
We define \( \mathbf{p_i} \in \mathbb{R}^{d_2} \) as the \( d_2 \)-dimensional sentence embedding corresponding to post \( p_i \). Similarly, \( \mathbf{c_j(p_i)} \in \mathbb{R}^{d_2} \) is the \( d_2 \)-dimensional sentence embedding corresponding to comment \( c_j(p_i) \). Let \( \mathbf{f_{p_i}} \) and \( \mathbf{f_{c_j(p_i)}} \) be the representations for post \( p_i \) and comment \( c_j(p_i) \), respectively, which aggregate both the textual content and user behavior features. These features are defined as follows.
\small
\begin{equation}
\label{equa1}
 \mathbf{f_{p_i}} = \mathbf{p_i} \, || \, \mathbf{u_p}
\end{equation}
\begin{equation}
\label{equa2}
 \mathbf{f_{c_j(p_i)}} = \mathbf{c_j(p_i)} \, || \, \mathbf{u_c}
\end{equation}
\normalsize
where \( \mathbf{u_p} \) and \( \mathbf{u_c} \) represent the feature vectors for the users who created the post \( p_i \) and made the comment \( c_j(p_i) \), respectively. The symbol \( || \) is the concatenation operator.

We use the sentence transformer model \citep{reimers-2020-multilingual-sentence-bert} to represent each sentence in a 256 ($d_2$) dimensional representation. First, we use BERT to create 768 dimensional word embeddings.
%we create word embeddings of dimension 768 using the BERT model. 
Then, we combine these word embeddings into a sentence embedding using a pooling layer. Here, we apply a $\tanh$ activation function and use a dense layer with 256 output features (i.e., \(d_2\)).
%to generate the final representation of the sentence.
Now, we concatenate the text and user embeddings and use this 384-dimensional embedding to represent the nodes that are further fed to the GNN module.

\subsection{GNN module}
% TODO Need to expalin the models
We consider the following standard GNN models for our experiments. These are \textsc{GCN} \citep{kipf2017semi}, \textsc{GraphSage} \citep{hamilton2018inductiverepresentationlearninglarge} and \textsc{GAT} \citep{veličković2018graphattentionnetworks}. 
\if{0}
\begin{itemize}
    \item \textbf{GCN} \citep{kipf2017semi} is an efficient way for semi-supervised learning on graph-structured data, encoding both the local graph and node features. We consider multi-layer GCN. The layer-wise propagation rule is as follows.
    \begin{equation}
    H^{(l+1)} = \sigma \left( D^{-1/2} (A + I) D^{-1/2} H^{(l)} W^{(l)} \right)
    \end{equation}
    where \( H^{(l)} \) represents the node feature matrix at layer \( l \),  \( D_{ii}=\sum{j}{A_{ij}} \) is the degree matrix and \( I \) is the identity matrix, \( W^{(l)} \) is the trainable weight matrix at layer \( l \) and \( \sigma \) is the activation function.
    \item \textbf{GraphSage} \citep{hamilton2018inductiverepresentationlearninglarge} : 
    \item \textbf{GAT} \citep{veličković2018graphattentionnetworks} :
\end{itemize}\fi
For all these models we consider three layers $(l=3)$. The dimensions of each node’s hidden feature vectors are 64. The activation function used is \textsc{ReLU}. Now, we consider a graph $L_i$ for each data sample $d_i$, with nodes represented by $\mathbf{f_{p_i}}$, $\mathbf{f_{c_j(p_i)}}$ and $E_i$ representing the edges. The features obtained from the above module and the graph $L_i$ are fed to GNN to learn graph-level representations. Let \( \mathbf{l_{p_i}} \) and \( \mathbf{l_{c}} \) denote the embeddings of the post and comment nodes, respectively, obtained from the GNN. The final graph-level representation \( \mathbf{L_i} \) for the data sample $d_i$ is computed as shown in the equation \ref{equa3}.
\small
\begin{equation}
\label{equa3}
\mathbf{L_i} = \lambda * \mathbf{l_{p_i}} + (1 - \lambda) *\frac{1}{|C(p_i)|}\sum_{c \in C(p_i)} \mathbf{l_c} 
\end{equation}
\normalsize
$\lambda$ is obtained using hyperparameter optimization framework \texttt{optuna}\footnote{\url{https://optuna.org/}} by minimizing the validation loss.

\subsection{Fake news detection module}
The final representation \( \mathbf{L_i} \)  obtained for each data sample $d_i$ is then passed through a fully connected layer to classify the sample as either fake (F) or true (T). The model is trained using Adam optimizer with a learning rate of 0.01. The loss function utilized in our model is the cross-entropy loss.

\subsection{Cold user behavior mapper module.}
\begin{figure}[!ht]
    \centering
    
 \subfigure[\footnotesize \textcircled{1},\textcircled{2} represent post similarity and reaction similarity heuristics respectively. In this example, we consider $k_1=3$ and $k_2=4$. The cold user is represented by the average of user embeddings of train users represented in \textcolor{royalblue}{blue}.]{
        \includegraphics[height=7cm, width=0.40\textwidth]{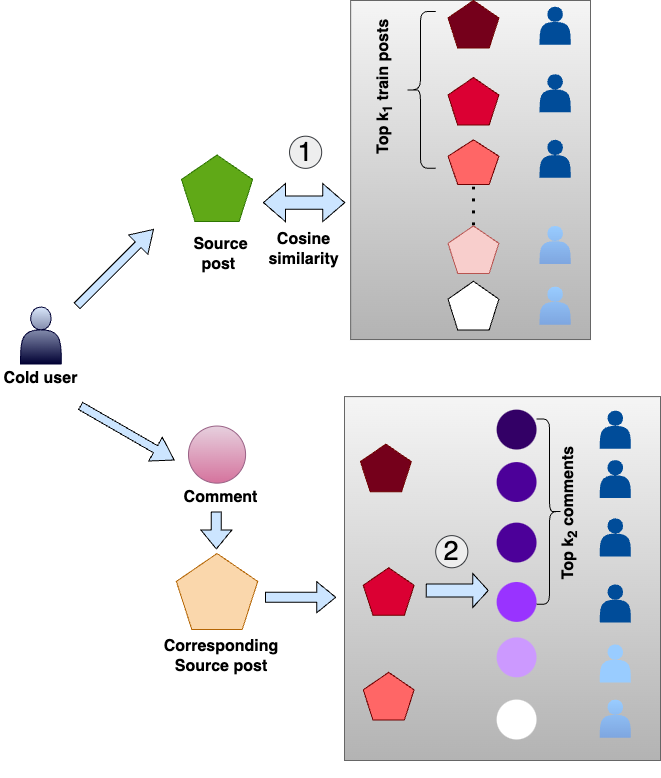}
        \label{fig:heur_m}
    }

    \hspace{0.05\textwidth}
      
    \subfigure[\footnotesize Representation of comments for comparison, following the \textcircled{3} historical reaction similarity. Each comment is represented as the sum of text embeddings from first-level comment to the current comment along the comment chain. ]{
        \includegraphics[height=7cm, width=0.40\textwidth]{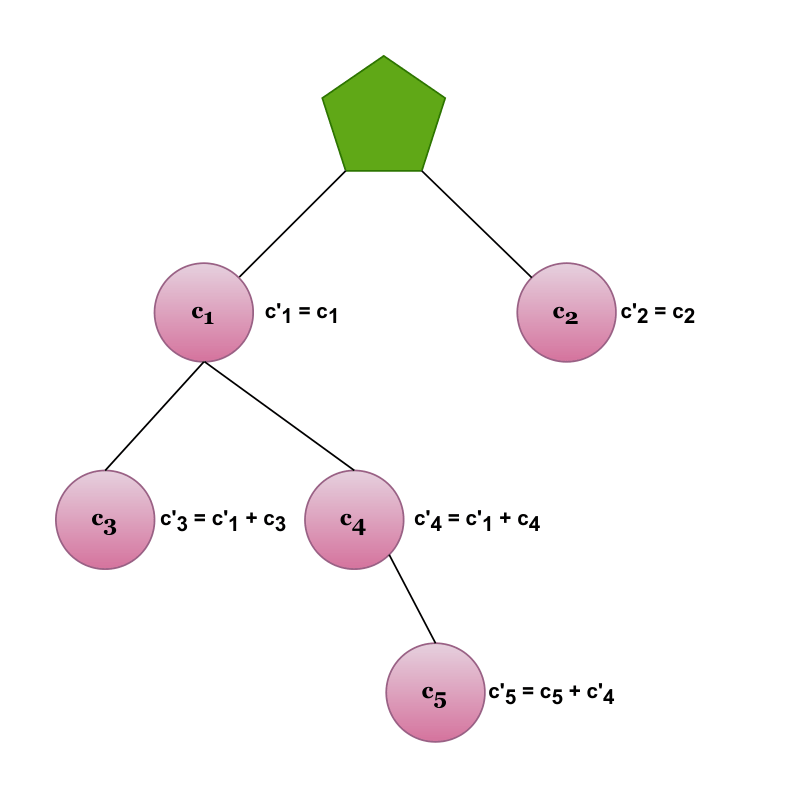}
        \label{fig:heur_rep}
    }
  
    \caption{Cold user behavior mapper module. }
    \label{fig:main_heur}
\end{figure}

Instead of random initialization, we propose a few heuristics to extract robust representation for cold users based on available data for more accurate approximation of the user behavior. \ksk{These heuristics are built upon the collaborative filtering approach used in the recommendation system (\citet{article_cf}).}
% The three heuristics that follow are intended to extract a robust representation for these users.

\begin{compactenum}
    \item \textbf{Post similarity}: Users with similar interest and preferences tend to post comparable content.
    
    \item \textbf{Reaction similarity}: Users often react similarly for similar content. This heuristic focuses on aligning user reactions to content.
    
    \item \textbf{Historical reaction similarity}: The way users react depends on the historical context of the comments chain. This takes into account the sequence and context of previous reactions to infer the user's current behavior.
\end{compactenum}

We apply the above three heuristics at different levels to find a representation to cold users. Figure \ref{fig:main_heur} represents our approach to represent the cold users. In this figure, dark colors are used for posts (comments) to represent higher similarity with the current post (comment), while light colors indicate lower similarity.
\begin{compactitem}
    \item \textbf{Representation of cold users at the source post}: We utilize the text embedding of the cold user's source post, then compute the cosine similarity with the text embeddings of posts from the training dataset commensurate with the first heuristic. Now, we identify the top $k_1$ posts which are most similar and represent the cold user with the average of user embeddings of $k_1$ train users who posted these most similar posts. This is illustrated in Figure \ref{fig:heur_m} where we consider $k_1=3$.
    \item \textbf {Representation of cold users who commented}: We first use the source post on which the comment of the cold user ($c_{cold}$) is made and find the top $k_1$ similar posts to this source post following the approach just mentioned above. We now collect all the comments to these $k_1$ posts (call this set $\mathcal{C}_{collect}$). We first represent $c_{cold}$ as well as each of the comment in $\mathcal{C}_{collect}$ as the sum of the text embeddings from the first level comment (i.e., the one on the post directly) to the current comment (i.e., $c_{cold}$ or one of the members of $\mathcal{C}_{collect}$) along the comment chain as shown in the Figure \ref{fig:heur_rep}, commensurate with the third heuristic. This transformation in $c_{cold}$ as well as the members of $\mathcal{C}_{collect}$ enables them to incorporate the characteristics of all the comments appearing before them in their respective chains. Now we compute the cosine similarity of the transformed $c_{cold}$ with each of the transformed member of $\mathcal{C}_{collect}$ and return the top $k_2$ similar comments which is commensurate with the second heuristic. Finally, we represent the cold user as the average of the user embeddings of $k_2$ train users who made those top $k_2$ similar comments. This is illustrated in Figure \ref{fig:heur_m} where we consider $k_2=4$.
    %Then, we consider the comments under these posts to map the similar comments following the second heuristic. We first represent the current comment of cold users and the comments considered from the training dataset as the sum of text embeddings from the first level comment to the current comment along the comment chain as shown in the figure \ref{fig:heur_rep}, following the third heuristic. Now, we calculate the cosine similarity to identify the top $k_2$ similar comments from the training dataset. Finally, we represent the cold user with the average of user embeddings of $k_2$ train users who commented similarly. .
\end{compactitem}
The values of $k_1$ and $k_2$ are obtained by optimizing the valid loss function. We used \texttt{faiss}\footnote{\url{https://github.com/facebookresearch/faiss}} for efficient similarity search. Note that, in this approach, the same user appearing multiple times in a particular data sample will be assigned different representations depending on the specific context of each occurrence.

\subsection{Comparision with other methods}
We choose three baselines for comparison with \uen. The first is the benchmark model proposed in the original paper introducing the Fakeddit dataset~\cite{nakamura2020rfakedditnewmultimodalbenchmark}. In addition, we compare \uen with a recent method proposed in~\cite{10.1145/3637528.3672024}, as well as with LLM-based baselines.
\\
\textbf{~\citet{nakamura2020rfakedditnewmultimodalbenchmark}}:  
To generate fixed-length BERT embedding vectors, the authors employ the bert-as-service tool~\cite{xiao2018bertservice} for converting the variable-length text or sentences into a 768-dimensional array for each Reddit submission title. In their experiments, they use the pre-trained \texttt{bert-large-uncased} model. They pass these embeddings through a trainable dense layer to obtain the final classification output. For our purpose, we reproduce the same setup.\\
\textbf{\citet{10.1145/3637528.3672024}}: We compare our {\uen} framework with \textbf{PSGT} -- Propagation Structure-Aware Graph Transformer \cite{10.1145/3637528.3672024}, -- a graph transformer based  approach~\cite{min2022transformergraphsoverviewarchitecture} that enhances reliability and interpretability by filtering out noisy information, focusing on task-relevant features, and simultaneously capturing long-range structural dependencies. The key difference of this model from graph transformer is that in place of multi-headed self attention, it uses graph-sampled multi-headed self attention. The authors employ three distinct graph masks -- a noise-filtered mask graph, designed to filter out noisy information among users; a structural positional encoding mask, aimed at capturing propagation depth and distance relationships between users; and a third mask focused on learning node interrelations while avoiding any structural biases. The input to the model has two parts -- the source news and the engaged users who either commented or replied to the news and their interaction patterns (similar to our input format). They use the pre-trained BERT model to extract the initial features from this input. 

\noindent \textbf{LLM bases baseline}: We prompt LLMs for fake news detection using the same approach mentioned in~\cite{10.1007/s11704-024-40674-6} to compare our model. Specifically, we use the instruction fine-tuned LLaMA 8B model checkpoint \texttt{meta-llama/Llama-3.1-8B-Instruct}, \ksk{ Qwen 8B model checkpoint \texttt{Qwen/Qwen3-8B}} and GPT-OSS 20B model checkpoint \texttt{openai/gpt-oss-20b} for our experiments. We consider two variants of the LLM: $\text{LLM}_{\text{cnt}}$, which takes only the source content as input, and $\text{LLM}_{\text{cnt+cmt}}$, which incorporates both the source content and associated comments.

%% file: Results.tex
\section{Results}
\label{sec:results}
\begin{table}[t]
    \centering
    \resizebox{0.75\columnwidth}{!}{
    \footnotesize
    \begin{tabular}{l|c|ccc}
        \toprule
        Dataset & Hyperparameters & \textsc{GCN} & \textsc{GraphSage} & \textsc{GAT} \\
        \midrule
        \multirow{3}{*}{Fakeddit} & $\lambda$ & 0.62  & 0.89  & 0.87  \\
        & $k_1$ & 19 & 28 & 11\\
        &$k_2$  & 72 & 96 & 51\\
        \midrule
        Gossipcop & k & 39 & 4 & 11 \\
        \bottomrule
    \end{tabular}
    }
    \caption{Hyperparameters used for each of the GNN models.}
    \label{tab:hyperparameters}
\end{table}

\renewcommand{\arraystretch}{1.5}
\begin{table}[t]
\footnotesize
    \centering
    \resizebox{0.95\columnwidth}{!}{ % Adjust the width to fit the column
    \begin{tabular}{l|ccccc}
        \toprule
        \textbf{Dataset} & \textbf{Method} & \textbf{Metrics} & \textbf{\textsc{GCN}} & \textbf{\textsc{GraphSage}} & \textbf{\textsc{GAT}} \\
        \midrule
        \multirow{6}{*}{Fakeddit} & \multirow{2}{*}{{\uen$_{w/o \text{\ user }}$}} & Accuracy & 0.77 & 0.81 & 0.79 \\ 
        & & Macro-F1 & 0.73 & 0.77 & 0.75 \\ 
        \cmidrule{2-6}
        & \multirow{2}{*}{{\uen$_{w/o \text{\ mapper }}$}} & Accuracy & 0.85 & 0.87 & 0.86 \\  
        & & Macro-F1 & 0.81 & 0.83 & 0.81 \\ 
        \cmidrule{2-6}
        & \multirow{2}{*}{{\uen}} & Accuracy & \cellcolor{green!30}\textbf{0.87} & \cellcolor{green!30}\textbf{0.87} & \cellcolor{green!30}\textbf{0.87} \\  
        & & Macro-F1 & \cellcolor{blue!30}\textbf{0.84} & \cellcolor{blue!30}\textbf{0.85} & \cellcolor{blue!30}\textbf{0.83} \\ 
        \midrule
        \multirow{6}{*}{Gossipcop} & \multirow{2}{*}{{\uen$_{w/o \text{\ user }}$}} & Accuracy & 0.89  & 0.91 & 0.92\\ 
        & & Macro-F1 & 0.89 & 0.91 & 0.92 \\ 
        \cmidrule{2-6}
        & \multirow{2}{*}{{\uen$_{w/o \text{\ mapper }}$}} & Accuracy & 0.92 & 0.94 & 0.95 \\  
        & & Macro-F1 & 0.92 & 0.94 & 0.95 \\ 
        \cmidrule{2-6}
        & \multirow{2}{*}{{\uen}} & Accuracy & \cellcolor{green!30}\textbf{0.92} & \cellcolor{green!30}\textbf{0.94} & \cellcolor{green!30}\textbf{0.95} \\  
        & & Macro-F1 &  \cellcolor{blue!30}\textbf{0.92} & \cellcolor{blue!30}\textbf{0.94} & \cellcolor{blue!30}\textbf{0.95} \\  
        \bottomrule
    \end{tabular}
    }
    \caption{\footnotesize Evaluation of the \uen frameworks. The variant with the best metric values are noted in \textbf{boldface}. Best accuracy is highlighted in \colorbox{green!30}{green} and best macro-F1 is highlighted in \colorbox{blue!30}{purple}.}
    \label{tab:overall_performance}
\end{table}
\subsection{Overall performance}
We evaluate the performance for the following variants of the framework using the three GNN models: \textsc{GCN}, \textsc{GraphSage}, and \textsc{GAT}. We report accuracy and a macro-averaged F1-Score.
\begin{compactitem}
 \item \(\mathbb{\uen}\): In this framework, we follow the same steps mentioned in the previous section.
 \item {$\mathbb{\uen}_{w/o \text{\ mapper }}$}: For the cold users, we initialize them with the average of all user embeddings in the \textit{global interaction-based graph}. Hence, in this framework, the cold user behavior mapper module is not utilized.
 \item {$\mathbb{\uen}_{w/o \text{\ user }}$}: Here, we ignore the user features and represent every node using only the textual content, i.e. $\mathbf{f_{p_i}} = \mathbf{p_i}$ and $\mathbf{f_{c_j(p_i)}} = \mathbf{c_j(p_i)}$
\end{compactitem}

Table \ref{tab:hyperparameters} displays the hyperparameters selected for our framework. The models have been trained on a NVIDIA A100 40GB GPU and implemented by \texttt{Pytorch}. Most models completed the training within a day. We trained the models for 20 epochs. Table \ref{tab:overall_performance} shows the performance of different variants of the proposed framework. 

\begin{table}[t]
    \centering
    \scriptsize
    \begin{tabular}{l||c|c}
        \toprule
        \textbf{Model} & \textbf{Fakeddit ($p$-value)} & \textbf{Gossipcop ($p$-value)} \\
        \midrule
        \textsc{GCN} & $1.8342 \times 10^{-7}$ & $2.1947\times 10^{-54}$ \\
        \textsc{GraphSage} & $7.0411 \times 10^{-96}$ & $3.284\times 10^{-19}$ \\
        \textsc{GAT} & $4.6118 \times 10^{-12}$ & $6.2534\times 10^{-90}$ \\
        \bottomrule
    \end{tabular}
    \caption{\footnotesize $p$-values comparing the prediction outcomes of {{\uen}$_{w/o \text{\ mapper }}$} with \uen using the Mann-Whitney U test.}
    \label{tab:p_value}
\end{table}

\subsection{Importance of user evidence}
Table~\ref{tab:overall_performance} clearly shows the advantage of including the user behavior based features ({{\uen}$_{w/o \text{\ mapper }}$}) over just using text based features ({{\uen}$_{w/o \text{\ user }}$}). For the fakeddit dataset, the accuracy improves by $8\%$, $6\%$ and $7\%$ for the models \textsc{GCN}, \textsc{GraphSage}, and \textsc{GAT}, respectively. Similarly, on the Gossipcop dataset, the inclusion of user evidence leads to a $3\%$ improvement in accuracy. Hence, the user evidence plays an important role in fake news detection. The complete \uen framework including the mapper module produces the best results across all the three GNN models and for both the metrics. Table~\ref{tab:p_value} shows the $p$-values comparing the prediction outcomes of {{\uen}$_{w/o \text{\ mapper }}$} with \uen using the Mann-Whitney U test. For all the GNN models the prediction outcomes are significantly different demonstrating that the macro-F1 improvements from {{\uen}$_{w/o \text{\ mapper }}$} to \uen are not by chance. 
%\ksk{Levene's test - suggestion for aaai. Should we do this?}

\if{0}To understand the importance of the user behavior, we compare the performance of the models with the inclusion and exclusion of user evidence.  shows the experimental results. We can observe that there is a significant difference in the performance between the frameworks.
\begin{itemize}
    \item The one with the \textbf{inclusion} of user evidence (corresponding to \(\mathbb{\uen_{w/o \text{\ mapper}}}\) ) 
    \item The one with the \textbf{exclusion} of user evidence (corresponding to \(\mathbb{\uen_{w/o \text{\ user}}}\) )
\end{itemize}\fi
\renewcommand{\arraystretch}{1.5}
\begin{table}[t]
\scriptsize
\centering
\begin{tabular}{l|rrr|rrr}
\toprule
\textbf{Category} & \multicolumn{3}{c|}{\textbf{Fakeddit}} & \multicolumn{3}{c}{\textbf{Gossipcop}} \\
 & \textbf{Total} & \textbf{True} & \textbf{Fake} & \textbf{Total} & \textbf{True} & \textbf{Fake} \\
\midrule
Total         & 116,639 & 86,443 & 30,196 & 1,094 & 557 & 537 \\
(0.5,1]       & 31,327  & 23,080 & 8,247  & 838   & 509 & 329 \\
(0,0.5]       & 63,915  & 50,758 & 13,157 & 240   & 44  & 196 \\
0             & 21,397  & 12,605 & 8,792  & 16    & 4   & 12 \\
\bottomrule
\end{tabular}
\caption{\footnotesize Distribution of test samples based on overlap ratio between the users in the sample and the \textit{global interaction-based graph.} Even the last row, representing the existence of solely \textit{cold users}, has a significant number of samples for Fakeddit.}
\label{table4}
\end{table}

\renewcommand{\arraystretch}{1.5} % Increases row height
\begin{table}[t]
\footnotesize
\resizebox{\columnwidth}{!}{%
\begin{tabular}{c||cc|cc|cc}
\toprule
 \multirow{2}{*}{\textbf{Category}} &                                   \multicolumn{2}{c|}{\textbf{\(\mathbb{\uen_{\textbf{w/o \ user}}}\)}} &
 \multicolumn{2}{c|}{\textbf{\(\mathbb{\uen_{\textbf{w/o \ mapper}}}\)}} & \multicolumn{2}{c}{\textbf{\uen}} 
 \\ 
 % \cline{3-8}
% \cmidrule(l{10pt}r{10pt}){3-4}
% \cmidrule(l{10pt}r{10pt}){5-6}
% \cmidrule(l{10pt}r{10pt}){7-8}                 
& \multicolumn{1}{c}{\textbf{Accuracy}} & \multicolumn{1}{c|}{\textbf{Macro-F1} } 
& \multicolumn{1}{c}{\textbf{Accuracy}} & \multicolumn{1}{c|}{\textbf{Macro-F1} } 
& \multicolumn{1}{c}{\textbf{Accuracy}} & \multicolumn{1}{c}{\textbf{Macro-F1} } \\ 
% \hline \hline
\midrule
% \multirow{3}{*}{\textsc{\textbf{GCN}}}   & 
\multicolumn{7}{c}{\textsc{\textbf{GCN}}} \\ \cmidrule(l{75pt}r{75pt}){1-7}
(0.5,1{]}                  
                       & \multicolumn{1}{c}{0.83}  & \multicolumn{1}{c|}{0.79}   
                       & \multicolumn{1}{c}{0.93} & \multicolumn{1}{c|}{0.91}    
                       & \multicolumn{1}{c}{0.93}   &  0.92 \\ %\cline{2-8}
(0,0.5{]}                  
                       & \multicolumn{1}{c}{0.78}     & \multicolumn{1}{c|}{0.72}
                       & \multicolumn{1}{c}{0.89} & \multicolumn{1}{c|}{0.84} 
                       & \multicolumn{1}{c}{0.89}   &  0.84 \\ %\cline{2-8}       
0                         
                       & \multicolumn{1}{c}{0.66}     & \multicolumn{1}{c|}{0.65}
                       & \multicolumn{1}{c}{0.63} & \multicolumn{1}{c|}{0.61} 
                       & \multicolumn{1}{c}{\cellcolor{green!30}\textbf{0.72}} 
                       &  \cellcolor{blue!30}\textbf{0.71} \\  
\midrule \midrule
\multicolumn{7}{c}{\textsc{\textbf{GraphSage}}} \\ 
\cmidrule(l{75pt}r{75pt}){1-7}
% \multirow{3}{*}{\textsc{\textbf{GraphSage}}}   & 
(0.5,1{]}                  
                       & \multicolumn{1}{c}{0.86}     & \multicolumn{1}{c|}{0.83}
                       & \multicolumn{1}{c}{0.93} & \multicolumn{1}{c|}{0.92}  
                       & \multicolumn{1}{c}{0.93}   &  0.91 \\ %\cline{2-8}
(0,0.5{]}                  
                       & \multicolumn{1}{c}{0.83}     & \multicolumn{1}{c|}{0.77} 
                       & \multicolumn{1}{c}{0.90} & \multicolumn{1}{c|}{0.86} 
                       & \multicolumn{1}{c}{0.88}   &  0.84 \\ %\cline{2-8}
0                         
                       & \multicolumn{1}{c}{0.69}     & \multicolumn{1}{c|}{0.68} 
                       & \multicolumn{1}{c}{0.69} & \multicolumn{1}{c|}{0.66} 
                       & \multicolumn{1}{c}{\cellcolor{green!30}\textbf{0.76}} 
                       &  \cellcolor{blue!30}\textbf{0.75} \\  
\midrule \midrule
\multicolumn{7}{c}{\textsc{\textbf{GAT}}} \\ 
\cmidrule(l{75pt}r{75pt}){1-7}
%\multirow{3}{*}{\textsc{\textbf{GAT}}}   & 
(0.5,1{]}                  
                       & \multicolumn{1}{c}{0.84}     & \multicolumn{1}{c|}{0.81}
                       & \multicolumn{1}{c}{0.93} & \multicolumn{1}{c|}{0.91}  
                       & \multicolumn{1}{c}{0.93}   &  0.90 \\ %\cline{2-8}
(0,0.5{]}                  
                       & \multicolumn{1}{c}{0.80}     & \multicolumn{1}{c|}{0.75}
                       & \multicolumn{1}{c}{0.89} & \multicolumn{1}{c|}{0.84}  
                       & \multicolumn{1}{c}{0.89}   &  0.83 \\ %\cline{2-8}
0                         
                       & \multicolumn{1}{c}{0.68}     & \multicolumn{1}{c|}{0.67}
                       & \multicolumn{1}{c}{0.65} & \multicolumn{1}{c|}{0.62}  
                       & \multicolumn{1}{c}{\cellcolor{green!30}\textbf{0.73}} 
                       &  \cellcolor{blue!30}\textbf{0.71} \\  
\bottomrule
\end{tabular}%
}
\caption{\footnotesize Results for \textbf{Fakeddit} dataset, based on overlap ratio based buckets. The variant with the best metric values for the zero overlap buckets are noted in \textbf{boldface}. Best accuracy for this bucket is highlighted in \colorbox{green!30}{green} and best macro-F1 for this bucket is highlighted in \colorbox{blue!30}{purple}.}
\label{table3}
\end{table}

\renewcommand{\arraystretch}{1.5} % Increases row height
\begin{table}[t]
\footnotesize
\resizebox{\columnwidth}{!}{%
\begin{tabular}{c||cc|cc|cc}
\toprule
 \multirow{2}{*}{\textbf{Category}} &                                   \multicolumn{2}{c|}{\textbf{\(\mathbb{\uen_{\textbf{w/o \ user}}}\)}} &
 \multicolumn{2}{c|}{\textbf{\(\mathbb{\uen_{\textbf{w/o \ mapper}}}\)}} & \multicolumn{2}{c}{\textbf{\uen}} 
 \\ 
 % \cline{3-8}
% \cmidrule(l{10pt}r{10pt}){3-4}
% \cmidrule(l{10pt}r{10pt}){5-6}
% \cmidrule(l{10pt}r{10pt}){7-8}                 
& \multicolumn{1}{c}{\textbf{Accuracy}} & \multicolumn{1}{c|}{\textbf{Macro-F1} } 
& \multicolumn{1}{c}{\textbf{Accuracy}} & \multicolumn{1}{c|}{\textbf{Macro-F1} } 
& \multicolumn{1}{c}{\textbf{Accuracy}} & \multicolumn{1}{c}{\textbf{Macro-F1} } \\ 
% \hline \hline
\midrule
% \multirow{3}{*}{\textsc{\textbf{GCN}}}   & 
\multicolumn{7}{c}{\textsc{\textbf{GCN}}} \\ \cmidrule(l{75pt}r{75pt}){1-7}
(0.5,1{]}                  
                       & \multicolumn{1}{c}{0.92}  & \multicolumn{1}{c|}{0.91}   
                       & \multicolumn{1}{c}{0.95} & \multicolumn{1}{c|}{0.94}    
                       & \multicolumn{1}{c}{0.95}   &  0.94 \\ %\cline{2-8}
(0,0.5{]}                  
                       & \multicolumn{1}{c}{0.80}     & \multicolumn{1}{c|}{0.70}
                       & \multicolumn{1}{c}{0.85} & \multicolumn{1}{c|}{0.78} 
                       & \multicolumn{1}{c}{\cellcolor{green!30}\textbf{0.85}}   &  {\cellcolor{blue!30}\textbf{0.79}} \\ %\cline{2-8}         
\midrule \midrule
\multicolumn{7}{c}{\textsc{\textbf{GraphSAGE}}} \\ 
\cmidrule(l{75pt}r{75pt}){1-7}
% \multirow{3}{*}{\textsc{\textbf{GraphSage}}}   & 
(0.5,5{]}                  
                       & \multicolumn{1}{c}{0.94}     & \multicolumn{1}{c|}{0.93}
                       & \multicolumn{1}{c}{0.96} & \multicolumn{1}{c|}{0.96}  
                       & \multicolumn{1}{c}{0.96}   &  0.96 \\ %\cline{2-8}
(0,0.5{]}                  
                       & \multicolumn{1}{c}{0.83}     & \multicolumn{1}{c|}{0.73} 
                       & \multicolumn{1}{c}{0.88} & \multicolumn{1}{c|}{0.75} 
                       & \multicolumn{1}{c}{\cellcolor{green!30}\textbf{0.90}}   &  {\cellcolor{blue!30}\textbf{0.81}} \\ %\cline{2-8}
 
\midrule \midrule
\multicolumn{7}{c}{\textsc{\textbf{GAT}}} \\ 
\cmidrule(l{75pt}r{75pt}){1-7}
%\multirow{3}{*}{\textsc{\textbf{GAT}}}   & 
(0.5,1{]}                   
                       & \multicolumn{1}{c}{0.94}     & \multicolumn{1}{c|}{0.94}
                       & \multicolumn{1}{c}{0.96} & \multicolumn{1}{c|}{0.96}  
                       & \multicolumn{1}{c}{0.96}   &  0.96 \\ %\cline{2-8}
(0,0.5{]}                  
                       & \multicolumn{1}{c}{0.86}     & \multicolumn{1}{c|}{0.76}
                       & \multicolumn{1}{c}{0.91} & \multicolumn{1}{c|}{0.85}  
                       & \multicolumn{1}{c}{\cellcolor{green!30}\textbf{0.93}}   &  {\cellcolor{blue!30}\textbf{0.87}} \\ %\cline{2-8}

\bottomrule
\end{tabular}%
}
\caption{\footnotesize Results for \textbf{Gossipcop} dataset, based on overlap ratio based buckets. The variant with the best metric values for the (0,0.5] category are noted in \textbf{boldface}. Best accuracy for this bucket is highlighted in \colorbox{green!30}{green} and best macro-F1 for this bucket is highlighted in \colorbox{blue!30}{purple}.}
\label{table3_Gossipcop}
\end{table}

\renewcommand{\arraystretch}{1.5} % Increases row height

\subsection{Investigation of cold users}
\ksk{Fakeddit dataset contains approximately 1.5 million users, of which 430,439 are cold users. Among all users, 11\% engage in fake posts per user (general), while cold users contribute 7\% engagement in fake posts per cold user. This demonstrates that cold users contribute a considerable volume to the overall fake post engagement. This observation highlights the importance of carefully assessing model performance in the presence of cold users. To better understand the challenge they pose,} we divide the test data into buckets based on the extent of overlap between the users in the test data and the \textit{global interaction-based graph}. In particular we compute the overlap ratio as the number of unique users in the test sample who also exist in the \textit{global interaction-based graph} to the total number of unique users in the test sample. Based on this value we divide the test data into three buckets as follows.
\begin{compactenum}
     \item \textbf{$(0.5,1]$}: Samples with a user overlap ratio greater than 50\%. This category represents scenarios where the model is relatively familiar with the majority of the users in the test dataset.

 \item \textbf{$(0,0.5]$}: Samples with a user overlap ratio less than 50\% but not zero. This category includes scenarios where the model has less prior knowledge of the users in the test dataset.

 \item \textbf{$(0)$}: Samples with no user overlap. This category represents the most challenging scenario, where the model encounters cold users who have no prior footprint on the platform. Analyzing this category is crucial for evaluating the model's robustness and its ability to generalize to completely new user data. For the Gossipcop dataset we omitted this set for our experiments as it has a very small number of instances.
\end{compactenum}

Table~\ref{table4} shows the distribution of samples in each category in the test dataset. We observe that there is a substantial number of cold users in the test dataset. The performance of the framework \uen$_{\text{w/o mapper}}$ across the three categories is reported in Table~\ref{table3} and Table~\ref{table3_Gossipcop}. We observe that the framework performs better in the first bucket, achieving $93\%$ accuracy irrespective of the GNN model used for the Fakeddit dataset, and $95-96\%$ for the Gossipcop dataset. In the second bucket, there is a slight drop in performance on Fakeddit, with accuracies of $89\%$, $90\%$, and $89\%$ for \textsc{GCN}, \textsc{GraphSage}, and \textsc{GAT}, respectively. For Gossipcop, the drop is more pronounced, with a reduction of approximately $5-10\%$ in accuracy and $11-21\%$ in macro-F1. For the third category (applicable only to Fakeddit, as the number of samples is insufficient in Gossipcop), we observe a significant decline in performance, with accuracy dropping by approximately $20-25\%$. This clearly demonstrates that the presence of cold users adversely affects model performance. However, this impact is mitigated when using the full \uen framework. Compared to the \uen$_{\text{w/o mapper}}$ variant, the accuracy improves by $9\%$, $7\%$, and $8\%$ for \textsc{GCN}, \textsc{GraphSage}, and \textsc{GAT}, respectively, on Fakeddit. The macro-F1 score also increases by $9-10\%$. Similarly, for the second category on the Gossipcop dataset, the full framework leads to a $2\%$ increase in accuracy and improvements of $1\%$, $6\%$, and $2\%$ in macro-F1 for \textsc{GCN}, \textsc{GraphSage}, and \textsc{GAT}, respectively. \\%\ksk{The ablation studies and sensitivity analysis of the parameters are discussed in Appendix~\ref{Ablation}.}
\noindent\textbf{Ablation study on heuristics used in the cold user mapper module}
\renewcommand{\arraystretch}{1.5} % Increases row height
\begin{table}[t]
\footnotesize
\resizebox{\columnwidth}{!}{%
\begin{tabular}{c||cc|cc|cc}

\toprule
 \multirow{2}{*}{\textbf{Category}} &                                   \multicolumn{2}{c|}{\textbf{\(\mathbb{\uen_{\textbf{H1}}}\)}} &
 \multicolumn{2}{c|}{\textbf{\(\mathbb{\uen_{\textbf{H1+H2}}}\)}} & \multicolumn{2}{c}{\textbf{\(\mathbb{\uen_{\textbf{H1+H2+H3}}}\)}} 
 \\ 
 % \cline{3-8}
% \cmidrule(l{10pt}r{10pt}){3-4}
% \cmidrule(l{10pt}r{10pt}){5-6}
% \cmidrule(l{10pt}r{10pt}){7-8}                 
& \multicolumn{1}{c}{\textbf{Accuracy}} & \multicolumn{1}{c|}{\textbf{Macro-F1} } 
& \multicolumn{1}{c}{\textbf{Accuracy}} & \multicolumn{1}{c|}{\textbf{Macro-F1} } 
& \multicolumn{1}{c}{\textbf{Accuracy}} & \multicolumn{1}{c}{\textbf{Macro-F1} } \\ 
% \hline \hline
\midrule
% \multirow{3}{*}{\textsc{\textbf{GCN}}}   & 
\multicolumn{7}{c}{\textsc{\textbf{GCN}}} \\ \cmidrule(l{75pt}r{75pt}){1-7}
Overall                  
                       & \multicolumn{1}{c}{0.86}  & \multicolumn{1}{c|}{0.82}   
                       & \multicolumn{1}{c}{0.86} & \multicolumn{1}{c|}{0.83}    
                       & \multicolumn{1}{c}{0.87}   &  0.84 \\ %\cline{2-8}
(0.5,1{]}                  
                       & \multicolumn{1}{c}{0.93}  & \multicolumn{1}{c|}{0.92}   
                       & \multicolumn{1}{c}{0.93} & \multicolumn{1}{c|}{0.92}    
                       & \multicolumn{1}{c}{0.93}   &  0.92 \\ %\cline{2-8}
(0,0.5{]}                  
                       & \multicolumn{1}{c}{0.88}     & \multicolumn{1}{c|}{0.85}
                       & \multicolumn{1}{c}{0.89} & \multicolumn{1}{c|}{0.83} 
                       & \multicolumn{1}{c}{0.89}   &  0.84 \\ %\cline{2-8}       
0                         
                       & \multicolumn{1}{c}{0.67}     & \multicolumn{1}{c|}{0.64}
                       & \multicolumn{1}{c}{0.71} & \multicolumn{1}{c|}{0.70} 
                       & \multicolumn{1}{c}{\cellcolor{green!30}\textbf{0.72}} 
                       &  \cellcolor{blue!30}\textbf{0.71} \\  
\midrule \midrule
\multicolumn{7}{c}{\textsc{\textbf{GraphSAGE}}} \\ 
\cmidrule(l{75pt}r{75pt}){1-7}
Overall                  
                       & \multicolumn{1}{c}{0.87}  & \multicolumn{1}{c|}{0.83}   
                       & \multicolumn{1}{c}{0.87} & \multicolumn{1}{c|}{0.83}    
                       & \multicolumn{1}{c}{0.87}   &  0.85 \\ %\cline{2-8}
% \multirow{3}{*}{\textsc{\textbf{GraphSage}}}   & 
(0.5,1{]}                  
                       & \multicolumn{1}{c}{0.93}     & \multicolumn{1}{c|}{0.91}
                       & \multicolumn{1}{c}{0.93} & \multicolumn{1}{c|}{0.91}  
                       & \multicolumn{1}{c}{0.93}   &  0.91 \\ %\cline{2-8}
(0,0.5{]}                  
                       & \multicolumn{1}{c}{0.89}     & \multicolumn{1}{c|}{0.84} 
                       & \multicolumn{1}{c}{0.88} & \multicolumn{1}{c|}{0.82} 
                       & \multicolumn{1}{c}{0.88}   &  0.84 \\ %\cline{2-8}
0                         
                       & \multicolumn{1}{c}{0.71}     & \multicolumn{1}{c|}{0.69} 
                       & \multicolumn{1}{c}{0.75} & \multicolumn{1}{c|}{0.74} 
                       & \multicolumn{1}{c}{\cellcolor{green!30}\textbf{0.76}} 
                       &  \cellcolor{blue!30}\textbf{0.75} \\  
\midrule \midrule
\multicolumn{7}{c}{\textsc{\textbf{GAT}}} \\ 
\cmidrule(l{75pt}r{75pt}){1-7}
Overall                  
                       & \multicolumn{1}{c}{0.86}  & \multicolumn{1}{c|}{0.82}   
                       & \multicolumn{1}{c}{0.87} & \multicolumn{1}{c|}{0.82}    
                       & \multicolumn{1}{c}{0.87}   &  0.83 \\ %\cline{2-8}
%\multirow{3}{*}{\textsc{\textbf{GAT}}}   & 
(0.5,1{]}                  
                       & \multicolumn{1}{c}{0.93}     & \multicolumn{1}{c|}{0.91}
                       & \multicolumn{1}{c}{0.93} & \multicolumn{1}{c|}{0.91}  
                       & \multicolumn{1}{c}{0.93}   &  0.90 \\ %\cline{2-8}
(0,0.5{]}                  
                       & \multicolumn{1}{c}{0.88}     & \multicolumn{1}{c|}{0.84}
                       & \multicolumn{1}{c}{0.88} & \multicolumn{1}{c|}{0.82}  
                       & \multicolumn{1}{c}{0.89}   &  0.83 \\ %\cline{2-8}
0                         
                       & \multicolumn{1}{c}{0.68}     & \multicolumn{1}{c|}{0.65}
                       & \multicolumn{1}{c}{0.71} & \multicolumn{1}{c|}{0.70}  
                       & \multicolumn{1}{c}{\cellcolor{green!30}\textbf{0.73}} 
                       &  \cellcolor{blue!30}\textbf{0.71} \\  
\bottomrule
\end{tabular}%
}
\caption{\footnotesize Results for \textbf{Fakeddit} dataset, based on overlap ratio based buckets. The variant with the best metric values for the (0) category are noted in \textbf{boldface}. Best accuracy for this bucket is highlighted in \colorbox{green!30}{green} and best macro-F1 for this bucket is highlighted in \colorbox{blue!30}{purple}.}
\label{table3_heuristics}
\end{table}

\noindent \textit{Fakeddit}: Table~\ref{table3_heuristics} presents the accuracy and macro-F1 scores obtained when different combinations of heuristics are used in the Cold User Mapper Module. The heuristics considered are: (H1) Post Similarity, (H2) Reaction Similarity, and (H3) Historical Reaction Similarity. \\
\noindent For each combination of heuristics, we report the best results achieved. As seen in the table, different combinations contribute variably to performance, and including all heuristics typically yields the highest scores.\\
\noindent \textit{Gossipcop}: Table~\ref{table4_heuristics} presents the accuracy and macro-F1 scores for two heuristics: (H1) Post Similarity and (H2) Reaction Similarity. In the case of the GossipCop dataset, each source news item is unique and accompanied by associated tweets and retweets. Therefore, H1 is not applicable, as there are no repeated posts to compare for similarity. For this reason, we have excluded H1 from the evaluation on Gossipcop and focused solely on H2. As shown in the table, H2 alone performs well on this dataset.

\begin{table}[t]
\footnotesize
\centering
\resizebox{0.75\columnwidth}{!}{%
\begin{tabular}{c||cc|cc}
\toprule
\multirow{2}{*}{\textbf{Category}} & 
\multicolumn{2}{c|}{\textbf{\(\mathbb{\uen_{\textbf{H1}}}\)}} & 
\multicolumn{2}{c}{\textbf{\(\mathbb{\uen_{\textbf{H2}}}\)}} \\ 
% \cmidrule(lr){2-3} \cmidrule(lr){4-5}
& \textbf{Accuracy} & \textbf{Macro-F1} & \textbf{Accuracy} & \textbf{Macro-F1} \\ 
\midrule

\multicolumn{5}{c}{\textsc{\textbf{GCN}}} \\ 
\cmidrule(lr){1-5}
Overall & 0.92 & 0.92 & 0.92 & 0.92 \\
(0.5,1] & 0.95 & 0.94 & 0.95 & 0.94 \\
(0,0.5] & 0.84 & 0.76 & {\cellcolor{green!30}\textbf{0.85}} & {\cellcolor{blue!30}\textbf{0.79}} \\

\midrule \midrule
\multicolumn{5}{c}{\textsc{\textbf{GraphSAGE}}} \\ 
\cmidrule(lr){1-5}
Overall & 0.94 & 0.93 & 0.94 & 0.94 \\
(0.5,1] & 0.96 & 0.96 & 0.96 & 0.96 \\
(0,0.5] & 0.88 & 0.79 & {\cellcolor{green!30}\textbf{0.90}} & {\cellcolor{blue!30}\textbf{0.81}} \\

\midrule \midrule
\multicolumn{5}{c}{\textsc{\textbf{GAT}}} \\ 
\cmidrule(lr){1-5}
Overall & 0.95 & 0.95 & 0.95 & 0.95 \\
(0.5,1] & 0.96 & 0.96 & 0.96 & 0.96 \\
(0,0.5] & 0.92 & 0.86 & {\cellcolor{green!30}\textbf{0.93}} & {\cellcolor{blue!30}\textbf{0.87}} \\

\bottomrule
\end{tabular}%
}
\caption{\footnotesize Results for \textbf{Gossipcop} dataset, based on overlap ratio based buckets. The variant with the best metric values for the (0,0.5] category are noted in \textbf{boldface}. Best accuracy for this bucket is highlighted in \colorbox{green!30}{green} and best macro-F1 for this bucket is highlighted in \colorbox{blue!30}{purple}.}
\label{table4_heuristics}
\end{table}

\begin{table}[h]
\centering
\footnotesize
\resizebox{0.75\columnwidth}{!}{%
\begin{tabular}{lccc}
\toprule
\textbf{Model} & \(k_1 = 5\) & \(k_1 = 10\) & \(k_1 = 15\) \\
\midrule
GCN        & 74.88 $\pm$ 0.33 & 76.54 $\pm$ 0.45 & 76.63 $\pm$ 0.42 \\
GraphSAGE  & 75.84 $\pm$ 0.48 & 77.02 $\pm$ 0.50 & 77.29 $\pm$ 0.53 \\
GAT        & 77.29 $\pm$ 0.25 & 77.89 $\pm$ 0.34 & 77.63 $\pm$ 0.29 \\
\bottomrule
\end{tabular}
}
\caption{Validation accuracy (\%) of different models for varying \(k_1\) values on the Fakeddit dataset. Values are reported as mean ± standard deviation.}
\label{tab:param_sensitivity}
\end{table}

\noindent\textbf{Parameter sensitivity}: \ksk{To evaluate the sensitivity of our models to the key parameter \(k_1\), we computed the mean and standard deviation of the validation accuracy for each setting on the Fakeddit dataset. The results are summarized in Table~\ref{tab:param_sensitivity}.}

\begin{table*}[h]
\footnotesize
    \centering
    \resizebox{0.75\textwidth}{!}{ % Adjust the width to fit the column
    \begin{tabular}{l|c|c|ccc|c|cc|cc|cc|}
        \toprule
        \textbf{Dataset} & \textbf{Category} & \textbf{Metrics} & \multicolumn{3}{c|}{\textbf{\uen}} & \textbf{PSGT} & 
        \multicolumn{2}{c|}{\textbf{Llama}} &
        \multicolumn{2}{c|}{\textbf{Qwen}} &
         \multicolumn{2}{c|}{\textbf{GPT-OSS}}\\
        \cmidrule{4-6} \cmidrule{8-13} 
        & & & \textbf{\textsc{GCN}} & \textbf{\textsc{GraphSage}} & \textbf{\textsc{GAT}} & & \textbf{$\text{LLM}_{\text{cnt}}$} & \textbf{$\text{LLM}_{\text{cnt+cmt}}$} & \textbf{$\text{LLM}_{\text{cnt}}$} & \textbf{$\text{LLM}_{\text{cnt+cmt}}$} & \textbf{$\text{LLM}_{\text{cnt}}$} & \textbf{$\text{LLM}_{\text{cnt+cmt}}$} \\
        \midrule
        \multirow{7}{*}{Fakeddit} & \multirow{2}{*}{{Overall}} & Accuracy & \cellcolor{green!30}\textbf{0.87} & \cellcolor{green!30}\textbf{0.87} & \cellcolor{green!30}\textbf{0.87} & \cellcolor{green!30}\textbf{0.81} & \cellcolor{green!30}\textbf{0.63} & \cellcolor{green!30}\textbf{0.73}  & 
        0.60 &
        0.65 &
        0.61 & 0.64\\ 
        & & Macro-F1 & 0.84 & \cellcolor{blue!30}\textbf{0.85} & 0.83 & \cellcolor{blue!30}\textbf{0.76} & \cellcolor{blue!30}\textbf{0.52} & 0.54 & 
        0.51 & \cellcolor{blue!30}\textbf{0.55} &
        0.45 & 0.44\\ 
        \cmidrule{2-13}
        & \multirow{2}{*}{{(0.5,1{]}}} & Accuracy & 0.93 & 0.93 & 0.93 & 0.86 & 0.63 & 0.73 & 0.62 & 0.67&
        0.60 & 0.64\\ 
        & & Macro-F1 & 0.92 & 0.91 & 0.90 & 0.83 & 0.52 & 0.53&  0.55 & 0.59&
        0.44 & 0.43\\ 
        \cmidrule{2-13}
        & \multirow{2}{*}{{(0,0.5{]}}} & Accuracy & 0.89 & 0.88 & 0.89 & 0.83 & 0.68 & 0.79& 0.60 & 0.65 &
        0.65 & 0.69\\   
        & & Macro-F1 & 0.84 & 0.84 & 0.83 & 0.76 & 0.52 & 0.57& 0.52 & 0.57&
        0.46 & 0.45\\ 
        \cmidrule{2-13}
        & \multirow{2}{*}{{0}} & Accuracy & 0.72 & \cellcolor{green!30}\textbf{0.76} & 0.73  &\cellcolor{green!30}\textbf{0.67} & 0.54 & \cellcolor{green!30} \textbf{0.56} & \cellcolor{green!30} \textbf{0.55} & \cellcolor{green!30} \textbf{0.56}&
        0.51 & 0.51\\ 
        & & Macro-F1 & 0.71 & \cellcolor{blue!30}\textbf{0.75} & 0.71 & \cellcolor{blue!30}\textbf{0.64} & \cellcolor{blue!30} \textbf{0.48} & 0.50 & \cellcolor{blue!30} \textbf{0.48} &  \cellcolor{blue!30} \textbf{0.51}&
        0.42 & 0.39\\ 
        \midrule
        \multirow{7}{*}{Gossipcop} & \multirow{2}{*}{{Overall}} & Accuracy & 0.92 & 0.94 & \cellcolor{green!30}\textbf{0.95} & \cellcolor{green!30}\textbf{0.89} & - & - & - & -&
        - & -\\ 
        & & Macro-F1 & 0.92 & 0.94 & \cellcolor{blue!30}\textbf{0.95} & \cellcolor{blue!30}\textbf{0.89} & - & - & - & -&
        - & -\\ 
        \cmidrule{2-13}
        & \multirow{2}{*}{{(0.5,1{]}}} & Accuracy & 0.95 & 0.96 & 0.96 & 0.93 & - & - & - & -&
        - & -\\  
        & & Macro-F1 & 0.94 & 0.96 & 0.96 & 0.92 & - & - & - & -&
        - & -\\ 
        \cmidrule{2-13}
        & \multirow{2}{*}{{(0,0.5{]}}} & Accuracy & 0.85 & 0.90 & \cellcolor{green!30}\textbf{0.93} & \cellcolor{green!30}\textbf{0.80} & - & - & - & -& - & -\\ 
        & & Macro-F1 & 0.79 & 0.81 & \cellcolor{blue!30}\textbf{0.87} & \cellcolor{blue!30}\textbf{0.63} & - & - & - & -& - & -\\ 
        \bottomrule
    \end{tabular}
    }
    \caption{Comparison of results of \textbf{\uen} with other baselines. Best accuracy for each baseline is highlighted in \colorbox{green!30}{green} and best macro-F1 for this bucket is highlighted in \colorbox{blue!30}{purple}.}
    \label{tab:UEN_vs_PSGT}
\end{table*}
\ksk{From the table, we observe that all three models show stable performance across different \(k_1\) values, indicating that our results are robust with respect to this parameter.}

%We evaluate the approximate representation of users proposed in the Cold user behavior mapper module across the three categories to understand the workings of the module. For this, we compare the frameworks -  and $UEN$. We observe, especially for the last category where there is no overlap of users, 

% Please add the following required packages to your document preamble:
% \usepackage{multirow}

 %on the Fakeddit dataset, considering only the source content is $86.44\%$. We used the same BERT model according to the hyperparameters mentioned in the paper and trained on the Fakeddit dataset, but we consider the data splits based on ordering the timestamp of the source post. Now, the accuracy has dropped to $77\%$.

\subsection{Baseline results}
\textbf{\citet{nakamura2020rfakedditnewmultimodalbenchmark}}:We recompute the results on our dataset (see Table~\ref{table1}) using text-only features and the same hyperparameter settings as in their paper, and obtain an accuracy of $77\%$ for Fakeddit and $73\%$ for Gossipcop. For Fakeddit, this performance is comparable to that of {{\uen}$_{w/o \text{\ user }}$}, which also uses text-only features. However, for Gossipcop, the performance is worse.
\\
\textbf{\citet{10.1145/3637528.3672024}}:
We train the PSGT model for 20 epochs, initialising the source post and engaged users with the text features obtained in the Content and Behavior Representation module. Table \ref{tab:UEN_vs_PSGT} presents the results of the comparison between the \uen framework and PSGT. Our findings indicate that the \uen framework achieves a $6\%$ higher accuracy compared to PSGT for both the datasets. Notably, for the zero-overlap case, \uen reports $9\%$ improvement (\textsc{GraphSage}) over PSGT for the Fakeddit dataset and in $(0,0.5]$ case, \uen reports $13\%$ improvement (\textsc{GAT}) for the Gossipcop dataset.

\noindent \textbf{LLM-based baselines}: Both LLM variants perform worse than our model. Overall, for the Fakeddit dataset our model achieves $24\%$ and $14\%$ higher accuracy compared to the two LLM baselines, respectively. For zero-overlap cases, it outperforms the LLMs by \ksk{$21\%$} and $20\%$, respectively. We are unable to report results for the Gossipcop dataset, as the raw text is not available.

\subsection{Analysis of the results}
We analysed the performance of \uen further on \textit{zero overlap} cases in the Fakeddit dataset based on the size and depth of the propagation tree. We also conducted analyses based on token length and subreddit, but no clear patterns observed in those dimensions.

\noindent \textbf{Based on tree size}: 
% \begin{compactitem}
 The tree size (number of comments, reactions, and source posts) within the bucket of size $[1$--$10]$ accounts for $80\%$ of the data in this dataset. In this bucket, both PSGT and our model performed equivalently, achieving an accuracy of $78\%$. For the remaining $20\%$ of the samples, where the tree size is greater than 10, PSGT's accuracy dropped to the range of $14-17\%$, while our model (\textsc{GAT}) maintained an accuracy between $60-70\%$.
% \item \ksk{\textbf{Gossipcop}: For this dataset also $60\%$ of the data lies within the bucket of size $[1$--$30]$ , for which PSGT and our model performed equivalently. But for the remaining $40\%$ of the samples,  PSGT's accuracy dropped to the range of $40\%$, while our model maintained an accuracy above $80\%$.
% }
% \end{compactitem}

\noindent \textbf{Based on depth}:
% \begin{compactitem}
\ Depth-based analysis revealed that accuracy of PSGT model was $32\%$ at depth $3$, while our model achieved $68\%$. The drop in PSGT performance at depth 3 can be attributed to the fact that $74\%$ of the data at this depth had tree size greater than $10$.
% \item \ksk{\textbf{Gossipcop}: Almost $67\%$ of the data have depth 3,for which PSGT and our model performed equivalently with accuracy above $90\%$. The accuracy of PSGT model dropped for depth range $[7$--$9]$ as $90\%$ of the data at this depths have tree-sizes greater than $100$.}

% The most natural baseline for our paper is the work that introduced the Fakeddit dataset~\citep{nakamura2020rfakedditnewmultimodalbenchmark}. Note that this dataset contains both text and images together and performs a multimodal classification. The authors reported $86.44\%$ accuracy using the BERT model for the 2-way classification task on this original Fakeddit dataset using \textit{text only} features. We recompute the results for our dataset (see Table~\ref{table1}) using text only features and the same hyperparameter settings as their paper and obtain an accuracy of $77\%$. Not surprisingly, this is very similar to the performance of {{\uen}$_{w/o \text{\ user }}$} which also uses text only features.

%% file: Conclusion.tex
\section{Conclusion}
\label{sec:conclusion}
In this paper, we first show the importance of content from posts, comments, and user behavior from the user-user interactions in fake news detection task. Then, we showed that there is a significant drop in performance for real-world datasets due to the presence of cold users. Next, we proposed \uen framework capable of approximating the behavior of cold users. Our carefully crafted heuristics enables us to significantly improve the detection performance especially when the cold users have absolutely no footprint in the training data. Our approach is generic and can be easily extended to other similar datasets.